\documentclass[aps,prl,showpacs,twocolumn]{revtex4-1}

\usepackage{bm,amsmath,amssymb,amsfonts,hyperref}
\usepackage{graphicx,epsf,subfigure}
\usepackage{xspace}
\usepackage{textcomp}
\usepackage{color}
\usepackage{epstopdf}
\usepackage{standalone}
\usepackage{layouts}

\newif\ifhyper
\hypertrue
\ifhyper
\hypersetup{
  citecolor = {green},
  urlcolor = {blue} 
} 

\newcommand{\sect}[1]{\noindent {\it{#1} --\xspace}}
\newcommand{\fref}[1]{Fig.~\ref{#1}}

\newcommand{\ie}{{\it i.e.}\xspace}
\newcommand{\eg}{{\it e.g.}\xspace}

\begin{document}

\title{Kardar-Parisi-Zhang universality in the phase distributions of one-dimensional  exciton-polaritons}

\author{Davide Squizzato}
\affiliation{Univ. Grenoble Alpes and CNRS, LPMMC,  F-38000 Grenoble}
\author{L\'eonie Canet}
\affiliation{Univ. Grenoble Alpes and CNRS, LPMMC,  F-38000 Grenoble}
\author{Anna Minguzzi}
\affiliation{Univ. Grenoble Alpes and CNRS, LPMMC,  F-38000 Grenoble}

\begin{abstract}
Exciton-polaritons under driven-dissipative conditions exhibit a condensation 
transition which belongs to a different universality class than equilibrium Bose-Einstein condensates.
 By numerically solving  the generalized Gross-Pitaevskii equation with realistic experimental parameters, 
 we show that one-dimensional exciton-polaritons display fine features of Kardar-Parisi-Zhang (KPZ) dynamics.
  Beyond the scaling exponents, we show
  that  their phase distribution follows the Tracy-Widom form predicted 
  for KPZ growing interfaces.   We moreover evidence
   a crossover to the stationary Baik-Rains statistics.
    We finally show that these features are unaffected on a certain timescale
   by the presence of a smooth disorder often present in experimental setups.
\end{abstract}

\maketitle

Non-equilibrium systems exhibit a large variety of critical behaviors, some of which having no counter-parts in equilibrium systems.
  This is the case for generic scale invariance, or self-organized criticality, 
  which is realized for instance in the celebrated Kardar-Parisi-Zhang equation \cite{kardar1}.
  Originally derived to describe the kinetic roughening of growing interfaces, it arises in connection with an extremely
 large class of non-equilibrium or disordered systems \cite{halpinhealy1,krug1} %, such as Burgers equation for randomly stirred fluids,
  %directed polymers in random media, dissipative transport, or flux lines in superconductors, turbulent liquid crystals.
   and  has therefore become a paradigmatic model in physics for non-equilibrium scaling and phase transitions,
  and    an archetype in mathematics of  stochastic processes with non-Gaussian statistics \cite{corwin1}. 
  
 Recently,  KPZ dynamics has been unveiled in a condensed matter system, the exciton-polariton  condensate \cite{ji1,gladilin1,he1,altman1, wouters2,sieberer1}.
 Exciton-polaritons (EP) are elementary excitations arising in a semiconductor microcavity coupled to cavity photons. Since EP have a finite lifetime, a stationary state is obtained under pumping. In these driven-dissipative conditions,
  the EP gas exhibits a non-equilibrium Bose-Einstein condensation transition, whose properties are intensively investigated \cite{carusotto1}.
  It was shown that in a certain regime, the dynamics of the phase of the condensate can be mapped onto a KPZ equation.
  Whereas this finding was confirmed numerically in one dimension (1D) for well-chosen parameters \cite{he1}, it was also suggested that this regime may not be accessible  in current experimental systems.

 In this paper, we re-examine the experimental realizability of KPZ universality in the 1D EP condensate.
  For this, we accurately model the system, taking into account realistic momentum-dependent losses and the quartic part of the dispersion of the polaritons. From this model, with actual experimental parameters, we numerically demonstrate that  KPZ physics is observable under current  experimental conditions. Moreover, beyond the KPZ scaling, we show that advanced KPZ properties can be 
  accessed in this system, considering in particular the statistics of the phase fluctuations.

 Indeed, a breakthrough in the understanding of the statistical properties of a 1D KPZ interface was achieved since 2010, with the derivation of the exact distributions of the fluctuations of the height of the interface, followed by other exact results for the two-point  correlations \cite{corwin1}.
 These advances highlighted remarkable features of the KPZ universality class: an unexpected connection with random matrix theory, with the appearance
 of Tracy-Widom (TW) distributions \cite{tracy-widom}, which are the distributions of the largest eigenvalues of matrices in the gaussian orthogonal (GOE) or unitary ensemble (GUE). It was shown that the interface is sensitive to the global geometry, or equivalently to the initial conditions, defining three sub-classes differing by their statistics: TW-GOE for flat \cite{calabrese2,calabrese3}, TW-GUE for  sharp-wedge (\ie delta-like) \cite{amir1,sasamoto1,calabrese1}, or Baik-Rains (BR) distribution  for stationary (\ie Brownian)   \cite{imamura1,imamura2} initial conditions, while sharing the same KPZ scaling exponents. 
  
This geometry-dependent universal  behavior was first observed
   experimentally in turbulent liquid crystal  \citep{takeuchi1,takeuchi2}. However, despite recent progress, experimental observations of KPZ universality are still scarce \cite{wakita1,maunuksela1}. We show that the EP system stands as a promising candidate.
 We numerically study the fluctuations of the phase of the EP condensate, and show that they precisely follow a TW-GOE distribution. Moreover, we show that a crossover between the TW-GOE distribution and the BR one can be observed.
  This crossover is expected in a finite-size system when the correlations become comparable with the system
    size but before finite-size effects begin to dominate, and is very difficult to access \cite{takeuchi3}.
    We show clear signatures of this crossover in the numerical distributions.
   Finally,  we investigate the effect of disorder, which is unavoidable in experimental systems, and
   show that KPZ physics is unaltered on a timescale related to the typical lengthscale of the disorder.

\sect{Model} 
A mean-field description of EP under incoherent pumping was introduced in \cite{wouters1}.
 In this description, the dynamics of the polariton condensate wavefunction  $\phi$ is given by
\begin{equation}\label{eq:phenomenep}
i\partial_t\phi=\left[ \mathcal{F}^{-1}\right.\left.[E_{LP}(k)](x)+\frac{i}{2}\left( Rn_r-\gamma_l \right)
+g|\phi^2| \right]\phi \, ,
\end{equation}
where the polaritonic reservoir density $n_r$ is determined by the rate equation
\begin{equation}
\partial_tn_r = P-\gamma_rn_r-Rn_r|\phi|^2\, .
\end{equation}
$E_{LP}(k)$ is the lower-polariton dispersion in  momentum space, $\mathcal{F}^{-1}$ denoting
 the inverse Fourier transform, $P$ is the pump, $R$ the amplification term, $\gamma_l$ 
 the polariton loss rate, $g$  the polariton-polariton interaction strength,  and $\gamma_r$  the reservoir loss rate.
In the phenomenological model \eqref{eq:phenomenep}, the dispersion is usually approximated 
 by a parabola of effective mass $m_{LP}$ with momentum-independent loss-rate $\gamma_l$.
 In this work, in order to describe more accurately the experiments, we also include quartic corrections 
 to this dispersion~\cite{carusotto1} $E_{LP}(k)\simeq \frac{\hbar^2k^2}{2m_{LP}}-\frac{1}{2\Omega}\left(\frac{\hbar^2k^2}{2m_{LP}}\right)^2$,
together with a momentum-dependent loss-rate $\gamma(k)=\gamma_l+\gamma^{(2)}k^2$, which originates from
 localized exciton reservoirs. These two effects result
 in an imaginary diffusion coefficient in the dynamics of the condensate, whose presence turns out to be crucial (see below).

In the case where the time scales in the reservoir and in the condensate are well separated,
  one can solve the dynamics of the reservoir density to obtain an effective equation for the polaritonic condensate,
   which in dimensionless form reads
\begin{align}\label{eq:ggpeplain}
i\partial_t\phi=\left[-(\right.&\left.1-iK_d)\nabla^2-K_c^{(2)}\nabla^4-(r_c-ir_d)+\right. \nonumber \\
&\left.(u_c- iu_d)|\phi|^2\right]\phi 
+\sqrt{\sigma}\xi \, ,
\end{align}
where we rescaled the time in units of $\bar{t}=\gamma_l^{-1}$, the space in units 
of $\bar{x}=(\hbar/2m_{LP}\gamma_l)^{1/2}$ and the condensate wavefunction in units 
of $\bar{\phi}=(\gamma_r (p-1)/Rp)^{1/2}$ with $p=PR/(\gamma_l\gamma_r)$. The parameters in \eqref{eq:ggpeplain} are 
related to the parameters of the microscopic model {\it via} $r_d=u_d=(p-1)/2$,
 $u_c=\gamma_r g (p-1)/(R \gamma_l p)$, $\sigma=Rp(p+1)/(2x^*\gamma_r(p-1))$,
  and $r_c$ determined from the stationary-homogeneous solution of \eqref{eq:ggpeplain}. The stochastic noise $\xi(x,t)$,
   with $\langle \xi(x,t)\xi^*(x',t')\rangle=2\delta(x-x')\delta(t-t')$ originates
    from the driven-disspative nature of the fluid \cite{ciuti1,wouters3}. 
     Equation \eqref{eq:ggpeplain} is a generalized Gross-Pitaevskii Equation (gGPE) with complex coefficients.

\sect{KPZ mapping} As shown in \cite{sieberer1}, by expressing the condensate wavefunction in a density-phase representation 
$\phi(x,t)=\sqrt{\rho(x,t)}\exp(i\theta(x,t))$ 
and performing a mean-field approximation over the density $\rho$ at the level of the Keldysh action for the
 EP, one obtains that the dynamics of the phase field $\theta$ is ruled by the KPZ equation
\begin{equation}\label{eq:kpzeq}
\partial_t\theta=\nu\nabla^2\theta+\frac{\lambda}{2}(\nabla\theta)^2+\sqrt{D} \eta\, ,
\end{equation}
where $\eta$ is a white noise with $\langle\eta(x,t)\eta(x',t')\rangle=2\delta(x-x')\delta(t-t')$ and 
$\nu=(K_cu_c/u_d+K_d),\,\lambda=-2(K_c-K_du_c/u_d), \, D=\sigma u_d\left(1+u_c^2/u_d^2 \right)/2r_d$ \citep{altman1}.
The original KPZ equation describes the dynamics of the height 
 of  a stochastically growing  interface. A 1D interface
  always roughens: it generically becomes scale-invariant. 
 Its profile
  can be characterized by the roughness, defined in term of $\theta$ as $w^2(L,t)=\langle \theta^2(x,t)-\langle \theta(x,t)\rangle_x ^2\rangle_{\xi,x}
$, where $\langle \cdot \rangle_x=1/L\int_x\cdot$ is the spatial average and 
$\langle \cdot \rangle_\xi$ the average over different realizations of the noise.
 The KPZ roughness is known to endow the Family-Vicsek scaling form \cite{family1,halpinhealy1}
\begin{equation} \label{eq:FW}
w(L,t) \sim t^\beta F(Lt^{-1/z})\sim  \left\{
  \begin{matrix}
    t^{\beta},& t<T_s\\
    L^{\chi},&  t>T_s
  \end{matrix}
\right.
\end{equation}
with $T_s\sim L^z$, and where the critical exponents take the exact values $\chi=1/2$ and  $z=3/2$,
$\beta =\chi/z=1/3$
 for the 1D KPZ universality class.

\sect{Numerical simulations}
 We numerically integrate the gGPE  \eqref{eq:ggpeplain}  using standard 
 Monte Carlo sampling of the noise \cite{werner1,dennis1}.
  The parameters in this equation
  depend on the material.  We use values typical for CdTe, used \eg~in Grenoble experiments:
  $m_{LP}=4\times10^{-5}m_e$, $\gamma_l=0.5$ ps$^{-1}$, $g=7.59\,10^2$ms$^{-1}$, $\gamma_r=0.02$ ps$^{-1}$, $R=400$ms$^{-1}$,
   $p=1.6$,  $K_d=0.45$ and $K_c^{(2)}=2.5 \times 10^{-3}$.
   In each simulation, we determine the wavefunction $\phi(t,x)$, and extract its phase $\theta(t,x)$. 
   We work in the low-noise regime, where the density fluctuations are negligible and topological defects  are absent. 
   In this regime the phase can be uniquely unwinded  to obtain $\theta \in (-\infty,\infty)$.

\begin{figure}[t]
\includegraphics[scale=1]{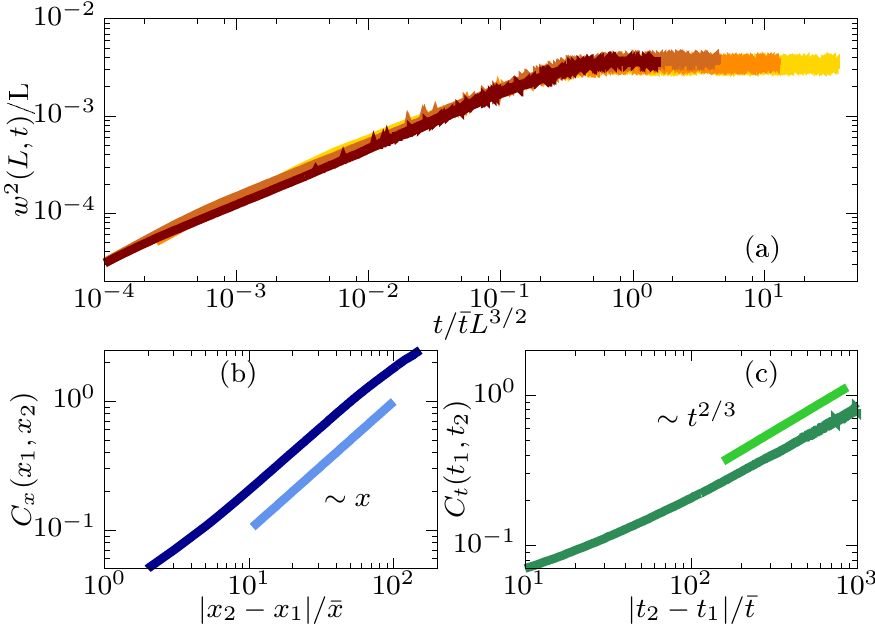}
\caption{(Color online)(a) Collapse of the roughness $w(L,t)$ with KPZ scaling $\chi=1/2$, $z=3/2$ for 
different system sizes $L/\bar{x}=2^7,2^8,2^9,2^{10}$ (size growing from lighter- to darker-color). Two-point (b) 
spatial and (c) temporal correlation together with the KPZ theoretical behavior, corresponding to a (stretched) exponential in $\rho_1(x_1,t,x_2,t)=e^{-C_x(x_1,x_2)}$ and $\rho_1(x,t_1,x,t_2)=e^{-C_t(t_1,t_2)}$}
\label{fig:kpzscalplain}
\end{figure}

\sect{KPZ scaling}   
  We have computed the roughness function  $w^2(L,t)$
   for the unwinded phase $\theta$  of EP.
   Our results are reported in \fref{fig:kpzscalplain},
    and show that using the KPZ exponents, 
    one obtains for the roughness a perfect collapse onto the expected Family-Vicsek scaling form \eqref{eq:FW}. This shows that the findings of \cite{he1} obtained for appropriate parameters can also be achieved with realistic experimental conditions.
    Let us stress that the inclusion of a momentum-dependent damping rate is crucial since it stabilizes the solution  for our choice of experimental parameters \footnote{Note that a similar imaginary diffusion constant was introduced in·\cite{he1} on an argument of relevance in the renormalization group sense, to stabilize the simulation.}, which corresponds to a  larger KPZ effective non-linearity parameter $|g|\equiv|\lambda|(D/2\nu^3)^{1/2}\simeq 0.48$ than that used  in \citep{he1}. This difference in parameters
    has an important  effect, since we obtain  KPZ scaling not only in the long-time saturation regime $t> T_{s}$, where 
    the roughness function reaches a constant in time, but also in the growth regime. 
    If the nonlinearity is too weak, only the initial Edward-Wilkinson (EW) scaling (with $\chi=1/2$ and $z=2$) is visible,
     and the crossover to the KPZ one cannot occur before $t$ reaches $T_{s}$.

  The KPZ scaling exponents can equivalently be determined directly from the condensate wavefunction, through
 the first-order correlation function, $\rho_1(x,t;x',t') = \langle \phi^*(x,t) \phi(x',t')\rangle_\xi$.
    Focusing on purely
   spatial or purely temporal correlations,  we define $C_x(x_1,x_2)|_t= - \log|\rho_1(x_1,t;x_2,t)|$ and
    $C_t(t_1,t_2)|_x=-\log |\rho_1(x,t_1;x,t_2)|$. At large distances, the main contribution to the correlation
     functions comes from the phase-phase correlations, that according to KPZ scaling should behave as
$\langle \theta(x_1,t) \theta(x_2,t) \rangle \sim |x_2-x_1|^{2\chi}$ and $\langle \theta(x,t_1) \theta(x,t_2) \rangle \sim |t_2-t_1|^{2\beta}$ respectively.
Our results for the correlation functions in the saturated regime $t>T_{s}$,  obtained from the numerical solution
 of the gGPE,  are shown in the lower panels of \fref{fig:kpzscalplain}. For both time and space correlations,
  we obtain $\chi\simeq0.49$ and $\beta\simeq0.31$ for $L/\bar{x}=2^{10}$ in close agreement with the KPZ exponents, which confirms our result from the roughness and validates our 
   phase reconstruction procedure. Let us note that the time-correlation is mandatory
    to discriminate between the KPZ and EW scalings, since only the
    $\beta$ exponent differs between the two in 1D.
   Quite remarkably, both space and time correlations are 
 routinely experimentally accessible (see \eg \cite{trichet1}).

\sect{Beyond scaling: Tracy-Widom statistics}
As emphasized in the introduction, 
  unprecedented theoretical advances have yielded
 the exact probability distribution of the fluctuations of the interface for sharp-wedge \cite{amir1,sasamoto1,calabrese1}, 
 flat \cite{calabrese2,calabrese3}, and stationary \cite{imamura1,imamura2} initial conditions.
  It was shown that at long times, the interface height $h$ behaves as  $h(x,t)\simeq v_\infty t +(\Gamma t)^{1/3}\chi(x,t)$ with  $\Gamma$ and $v_\infty$  non-universal parameters, and $\chi$ a random variable whose distribution is non-gaussian, and 
    exaclty given by the TW-GUE, TW-GOE or BR  distribution respectively.
    To further assess KPZ universality in EP systems, we thus study the fluctuations of the unwinded phase $\delta \theta= (\theta-\langle\theta\rangle_{\xi,x})$ of the condensate (which substracts the $v_\infty t$ term).
  In practice,  we use the gGPE simulations to determine  the distribution of the random variable $\tilde\theta(x,t)=\delta \theta /(\Gamma t)^{1/3}$, where 
    the parameter  $\Gamma$ is extracted
  from the numerical data using the relation 
   $\Gamma=\lim_{t\rightarrow \infty}\left(\langle\delta\theta^2(x,t)\rangle/\textrm{Var}_{\chi}\right)^{3/2}/t$ 
   \citep{halpinhealy2,takeuchi2}, with $\langle \delta\theta^2 \rangle=\langle (\theta-\langle\theta\rangle_{\xi,x})^2 \rangle_{\xi,x}$ and Var$_{\chi}$ is the theoretical value of the variance of the distribution. Note that since $\lambda<0$ for the EP system, the obtained distributions for $\tilde \theta$ are related to $P_{-\chi}$.

\begin{figure}[t]
\includegraphics{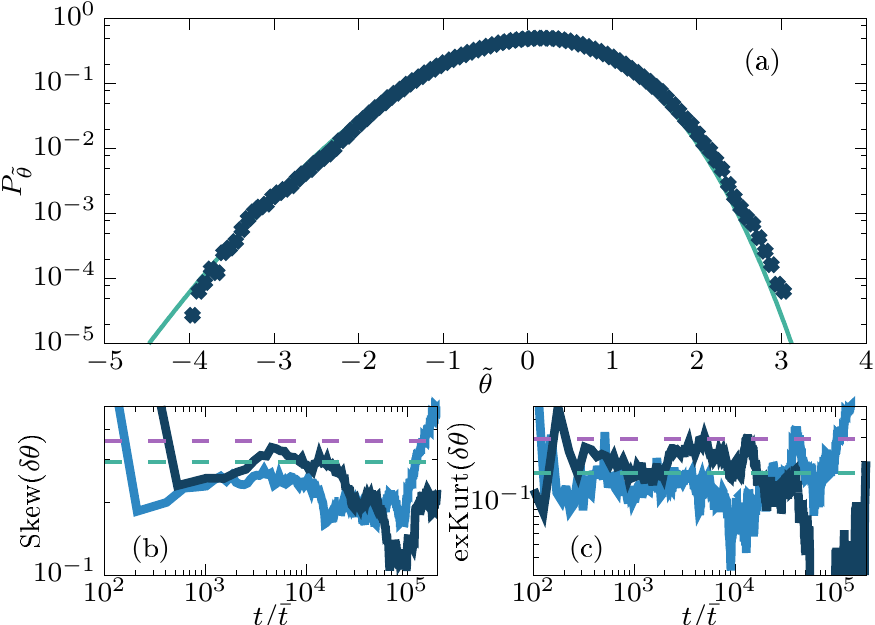}
\caption{(a) Distribution of 
%$\delta\theta(x=0,t)/(\Gamma t/\bar{t})^{1/3}$
$\tilde{\theta}$ for
 $L/\bar{x}=2^{10}$, together with the theoretical centered GOE-TW distribution. (b) Skewness  and (c) excess Kurtosis
  (lower panel) of the phase-field in the condensate for $L/\bar{x}=2^9,2^{10}$,
   together with the theoretical values for TW-GOE and 
   BR statistics (green and purple dashed-line respectively).  TW-GOE values are reached on a plateau around times  $t/\bar{t}\simeq 10^4$.}
\label{fig:cumulsplain}
\end{figure}

To gain insight on the nature of this distribution, we first compute universal ratios of 
    cumulants of $\delta\theta$, which do not depend on 
   $\Gamma$, namely the skewness and the excess kurtosis, defined 
 as $\textrm{Skew}(\delta\theta)=\langle \delta \theta^3\rangle/\langle \delta \theta^2\rangle^{3/2}$ 
 and $\textrm{eKurt}(\delta\theta)=\langle \delta \theta^4\rangle/\langle \delta \theta^2\rangle^{2}-3$ 
 respectively, with $\langle \delta\theta^n \rangle=\langle (\theta- \langle
  \theta \rangle_{\xi,x})^n \rangle_{\xi,x}$. These quantities are exactly zero for a Gaussian 
  distribution and are known numerically at arbitrary precision for the distributions associated with the 1D KPZ equation  \cite{prahofer1,bornemann1}.
  We find that they reach stationary values on plateaus depending on the system size but roughly 
 extending between  $t=10^3$ and $10^4$ in units of $\bar t$. The values of these plateaus are compatible with TW-GOE distribution (see \fref{fig:cumulsplain}-(b)).  
    We thus used the exact value of Var$_{{\tiny \rm TW-GOE}}$ to extract $\Gamma$, 
 and  recorded the probability  distribution of $\tilde\theta$ accumulated during the plateaus,
 which is represented in \fref{fig:cumulsplain}-(a).
  We find a close agreement with the theoretical TW-GOE curve.
   This result 
  provides a convincing confirmation that KPZ dynamics is relevant in EP systems. Note that the TW-GOE 
  distribution is associated  with flat (\ie spatially constant) initial conditions 
  for the $\theta$ field. This is  non trivial for the phase of EP, since a transcient non-universal
    regime exists before KPZ behavior sets in, that hinders which
  precise initial conditions are relevant for the associated KPZ equation.

 %  It's worth noting that they are different from just a generalisation of the roughness $w^2(L,t)$ to 
  % higher order $w^n(L,t)$ due to the fact that in the latter one subtracts the single-realization average
  %  $\langle \cdot \rangle_x$ instead of the average over the noise and thus is subjected to loose
  %   information about large deviation from the mean.

\sect{Beyond scaling: Baik-Rains statistics}
In a finite-size system, a crossover to the stationary KPZ dynamics, associated with the Baik-Rains distribution, is expected at sufficiently long times, but before finite-size effects dominate \cite{takeuchi2}.
 Indications of a change are manifest in \fref{fig:cumulsplain} since  the skewness 
  and excess Kurtosis depart from the plateaus at large times. However, this change is 
  hindered by the noise and finite-size effects, which become more and more relevant as the correlation length becomes
    comparable to the system size.

\begin{figure}[t]
\includegraphics{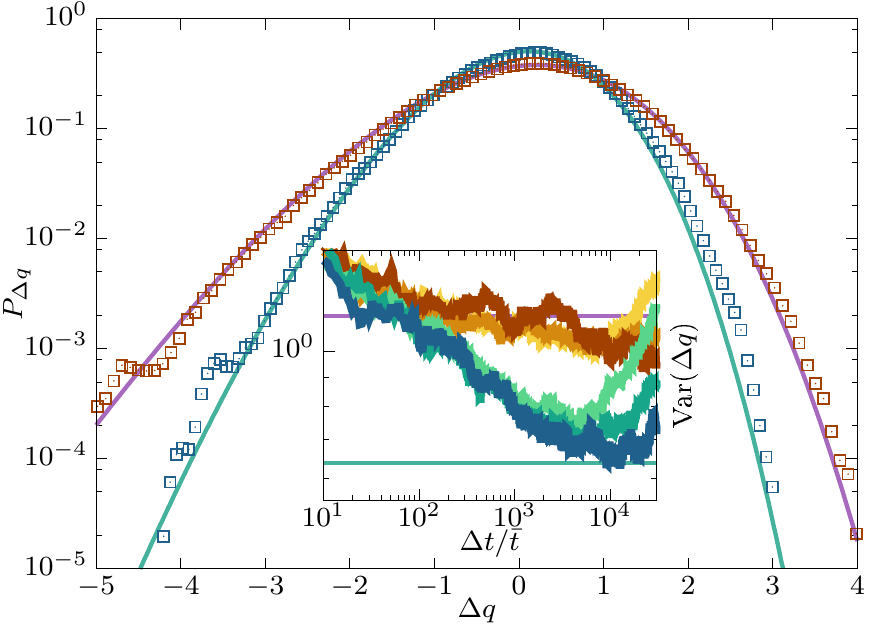}
\caption{(Color online) Distribution of $\Delta q(x,t_0,\Delta t)$  for  $\Delta t/t_0=10^3$ (light-green symbols)
 and $\Delta t/t_0=5\times 10^{-3}$ (purple symbols)  for $L/\bar{x}=2^{10}$, together with the theoretical  
    centered TW-GOE (green solid line)  and BR (purple solid line) distributions \cite{prahofer1}.  Inset (same color code): numerical and theoretical values of Var$(\Delta q)$ for different initial times $t_0/\bar t=10^2,2\times 10^4$ 
  (blue and brown color-scale respectively) and sizes $L/\bar x=2^8,2^9,2^{10}$ (increasing from lighter to darker).}
%The lower plateau is the one associated to TW-GOE statistics, reached for very large $\Delta t$. One can see that a higher plateau for  small $\Delta t$ appears increasing $t_0$ in large systems; This is consistent with the crossover to the stationary BR class, which is   further supported by the value of the ratio between the two plateaus, which is  $\simeq 1.781$, close to the theoretical value  Var$_{BR}/$Var$_{TW-GOE}=1.803$.}
\label{fig:twbrcross}
\end{figure}

In order to reduce finite-size effects and study  this crossover, we follow Takeuchi \citep{takeuchi2} and
  introduce a new variable $\Delta q(x,t_0,\Delta t)=(\delta\theta(x,t_0+\Delta t)-\delta \theta(x,t_0))/(\Gamma t)^{1/3}$.
  This variable is expected to display a  TW-GOE
   distribution for $t_0\rightarrow 0, \Delta t \rightarrow \infty$ and a BR distribution for 
   $t_0\rightarrow \infty, \Delta t \rightarrow 0$, with this precise   ordering of the limits. 
   The distribution of $\Delta q$ is plotted in \fref{fig:twbrcross} for different ratios $\Delta t/t_0$.
   Both TW and BR distributions are clearly identified. Our analysis hence shows that the homogeneous EP condensate is an ideal playground to observe 
 non-trivial out-of-equilibrium behavior associated with KPZ universality sub-classes.%with highly non-Gaussian distributions.

\begin{figure}
\includegraphics[scale=1]{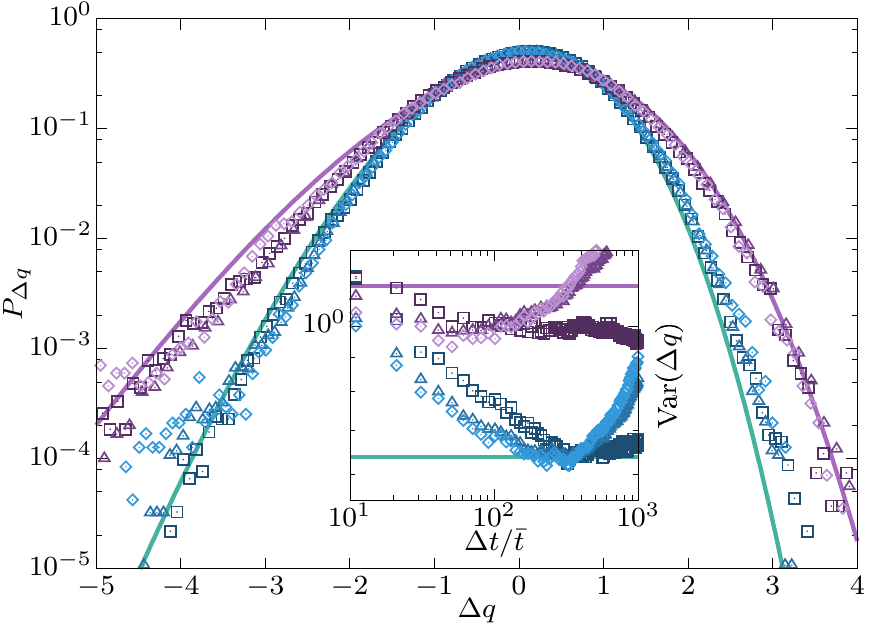}
\caption{(Color online)  Distribution of $\Delta q(x,t_0,\Delta t)$
 for different disorder correlation lengths $\ell_d/L=0.02,0.07,0.15$
  (from lighter to darker) for $\Delta t/t_0\simeq 10^2$ (blue symbols) and  $\Delta t/t_0\simeq 10^{-1}$
   (purple symbols), with $L/\bar{x}=2^{10}$. The solid lines correspond to centered TW-GOE (cyan)  and BR (violet)
    theoretical distributions  \cite{prahofer1}.  Inset (same parameters and color code): numerical
     and theoretical values for Var$(\Delta q)$.} 
\label{fig:dis}
\end{figure}

 \sect{Influence of the disorder}
In experimental setups,
  the inhomogeneities due to cavity imperfections give rise to a static disorder potential.
 We now investigate how this affects the KPZ physics.
  In order to model this disorder, we introduce a random-potential 
 $V_d(x)=|\mathcal{F}^{-1}[V_d(p)](x)|$ with $\left\langle V_d(x)V_d(x')\right\rangle= G(x-x')$, $V_d(p) 
=V_0 e^{ i \varphi} e^{-p^2\ell_d^2}$ and $\varphi$ a uniformly distributed random-variable in the range $[0,2\pi)$.
 By changing the correlation length $\ell_d$,  this describes any intermediate condition between a uniform  potential
  for  $\ell_d\rightarrow \infty$  to a white-noise disorder in the $\ell_d\rightarrow 0$ limit. We  focus 
  on  finite values $\ell_d$, corresponding to a smooth disorder  typical of  experiments.
 By means of the Keldysh formalism, one can show \cite{squizzato1}
that the inclusion of a static disorder in the Keldysh action leads to a non-local shift of the KPZ noise strength.
 %More precisely, the coefficient $D$ in the KPZ equation becomes $D_{dis}\tilde{\theta}(x,t)=D\tilde{\theta}(x,t)+\frac{1}{2}\int_{x'}G(x-x')\tilde{\theta}(x',t)$.
In the white-noise limit $\ell_d\rightarrow 0$ the presence of a disorder potential
 simply gives rise to a constant shift. However, in the case of a finite correlation length, the system 
 is characterized by an additional microscopic length scale, which affects the phase fluctuations for times longer than
  the  time scale   $t_d\sim \ell_d^{2/3}$. We hence expect that for $t<t_d$, KPZ physics is still observable, while
   for $t>t_d$ the features of the disorder become dominant. 
   The roughness and correlation functions computed in presence of the disorder indeed confirm this picture \cite{squizzato1}.
  We determined the distribution of the variable $\Delta q$, in both regimes of large and small $\Delta t/t_0$.
The results, presented in \fref{fig:dis}, show that for large $\Delta t/t_0$, the TW-GOE distribution
 is still accurately reproduced. Increasing $t_0$, the approach to the BR distribution is also clearly visible,  
 even if it  cannot be fully attained, since $t_0$ is limited to $t_d$ by the presence of the disorder.

\sect{Conclusion}
 We have shown that  universal KPZ physics can be observed  under realistic experimental conditions
   in the dynamics of the phase of one-dimensional
 exciton polaritons. %In particular, we have introduced in the model a 
% typical momentum-dependent damping and taken realistic experimental parameters for exciton-polaritons in semiconductor
%  microcavities.
 Our analysis shows that the roughness function, the first-order correlation functions and the
   probablity distributions of the phase fluctuations display universal KPZ features, such as a time stretched exponential decay of correlations
    and non-Gaussian probability distributions.  We evidenced in particular a crossover between TW-GOE and BR statistics, allowing 
 to probe two well-known sub-classes of 1D KPZ universality.
   Furthermore, we have shown that the presence of a static disorder 
   does not destroy  KPZ physics on sufficiently small timescales. This analysis suggests new experimental protocols for the observation of KPZ properties in exciton-polaritons. 
 %In outlook, one could further refine the model by including eg  the effect of thermal phonon fluctuations, 
 %as well as study the probability distributions in a higher-dimensional system, where ARE THEY KNOWN? 

\begin{acknowledgments}
The authors would like to thank Dario Ballarini, Denis Basko,   Maxime Richard and Alberto Rosso for
 inspiring discussions on both theoretical and experimental aspects.  
 This work is supported by French state funds ANR-10-LABX-51-01 (Labex LANEF du Programme d'Investissements d'Avenir) 
  and CNRS through grant  Infinity NEQ-DYN.
\end{acknowledgments}

\bibliographystyle{prsty}
%\bibliography{biblio}

\begin{thebibliography}{10}

\bibitem{kardar1}
M. Kardar, G. Parisi, and Y.-C. Zhang, Phys. Rev. Lett. {\bf 56},  889  (1986).

\bibitem{halpinhealy1}
T. Halpin-Healy and Y.-C. Zhang, Physics Reports {\bf 254},  215   (1995).

\bibitem{krug1}
J. Krug, Adv. Phys. {\bf 46},  139   (1997).

\bibitem{corwin1}
I. Corwin, Random Matrices: Theory and Applications {\bf 01},  1130001  (2012).

\bibitem{ji1}
K. Ji, V.~N. Gladilin, and M. Wouters, Phys. Rev. B {\bf 91},  045301  (2015).

\bibitem{gladilin1}
V.~N. Gladilin, K. Ji, and M. Wouters, Phys. Rev. A {\bf 90},  023615  (2014).

\bibitem{he1}
L. He, L.~M. Sieberer, E. Altman, and S. Diehl, Phys. Rev. B {\bf 92},  155307
  (2015).

\bibitem{altman1}
E. Altman, L. M. Sieberer, L. Chen, S. Diehl, and J. Toner, Phys. Rev. X {\bf 5},  011017  (2015).

\bibitem{wouters2}
M. Wouters and V. Savona, Phys. Rev. B {\bf 79},  165302  (2009).

\bibitem{sieberer1}
L.~M. Sieberer, M. Buchhold, and S. Diehl, Reports on Progress in Physics {\bf
  79},  096001  (2016).

\bibitem{carusotto1}
I. Carusotto and C. Ciuti, Rev. Mod. Phys. {\bf 85},  299  (2013).

\bibitem{tracy-widom}
C. Tracy and H. Widom, Commun. Math. Phys. {\bf 159},  151  (1994).

\bibitem{calabrese2}
P. Calabrese and P. Le~Doussal, Phys. Rev. Lett. {\bf 106},  250603  (2011).

\bibitem{calabrese3}
P. Calabrese and P. Le~Doussal, J. Stat. Mech.  P06001  (2012).

\bibitem{amir1}
G. Amir, I. Corwin, and J. Quastel, Commun. Pure Appl. Math. {\bf 64},  466
  (2011).

\bibitem{sasamoto1}
T. Sasamoto and H. Spohn, Nucl. Phys. B {\bf 834},  523   (2010).

\bibitem{calabrese1}
P. Calabrese, P.~L. Doussal, and A. Rosso, EPL (Europhysics Letters) {\bf 90},
  20002  (2010).

\bibitem{imamura1}
T. Imamura and T. Sasamoto, Phys. Rev. Lett. {\bf 108},  190603  (2012).

\bibitem{imamura2}
T. Imamura and T. Sasamoto, Journal of Statistical Physics {\bf 150},  908
  (2013).

\bibitem{takeuchi1}
K.~A. Takeuchi and M. Sano, Phys. Rev. Lett. {\bf 104},  230601  (2010).

\bibitem{takeuchi2}
K. Takeuchi and M. Sano, J. Stat. Phys. {\bf 147},  853  (2012).

\bibitem{wakita1}
J. ichi Wakita, H. Itoh, T. Matsuyama, and M. Matsushita, Journal of the
  Physical Society of Japan {\bf 66},  67  (1997).

\bibitem{maunuksela1}
J. Maunuksela, M. Myllys, O.-P. K\"ahk\"onen, J. Timonen, N. Provatas, M. J. Alava, and T. Ala-Nissila, Phys. Rev. Lett. {\bf 79},  1515  (1997).

\bibitem{takeuchi3}
K.~A. Takeuchi, Phys. Rev. Lett. {\bf 110},  210604  (2013).

\bibitem{wouters1}
M. Wouters and I. Carusotto, Phys. Rev. Lett. {\bf 99},  140402  (2007).

\bibitem{ciuti1}
C. Ciuti, G. Bastard, and I. Carusotto, Phys. Rev. B {\bf 72},  115303  (2005).

\bibitem{wouters3}
M. Wouters and V. Savona, Phys. Rev. B {\bf 79},  165302  (2009).

\bibitem{family1}
F. Family and T. Vicsek, Journal of Physics A: Mathematical and General {\bf
  18},  L75  (1985).

\bibitem{werner1}
M. Werner and P. Drummond, Journal of Computational Physics {\bf 132},  312
  (1997).

\bibitem{dennis1}
G.~R. Dennis, J.~J. Hope, and M.~T. Johnsson, Computer Physics Communications
  {\bf 184},  201   (2013).

\bibitem{Note1}
Note that a similar imaginary diffusion constant was introduced in·\cite {he1}
  on an argument of relevance in the renormalization group sense, to stabilize
  the simulation.

\bibitem{trichet1}
A. Trichet, E. Durupt, F. M\'edard, S.  Datta, A. Minguzzi, and M. Richard, Phys. Rev. B {\bf 88},  121407  (2013).

\bibitem{halpinhealy2}
T. Halpin-Healy and Y. Lin, Phys. Rev. E {\bf 89},  010103  (2014).

\bibitem{prahofer1}
M. Pr\"ahofer and H. Spohn, Phys. Rev. Lett. {\bf 84},  4882  (2000).

\bibitem{bornemann1}
F. Bornemann, Mathematics of Computation {\bf 79},  871  (2010).

\bibitem{squizzato1}
D. Squizzato, L. Canet, and A. Minguzzi, in preparation  (2017).

\end{thebibliography}

%\printinunitsof{in}\prntlen{\textwidth}

\end{document}